\documentclass[twocolumn,showpacs,preprintnumbers,showkeys,superscriptaddress]{revtex4}
\usepackage{tipa}
\usepackage{mathrsfs}
\usepackage{amssymb}
\usepackage{graphicx}
\usepackage{dcolumn}
\usepackage{bm}
\usepackage{graphicx}
\usepackage{float}%let figure float to where you want
\usepackage{longtable}
\usepackage{amsmath}

\newfont{\largemi}{cmmi10}
\newfont{\smallmi}{cmmi6}

\draft

\def\eqref#1{Eq.~(\ref{eq:#1})}

\begin{document}

\title{Nucleon pair approximation description of the low-lying structure of $^{108,109}$Te and $^{109}$I}

\author{H. Jiang \footnote{huijiang@shmtu.edu.cn}}
 \affiliation{School of Arts and
Sciences, Shanghai Maritime University, Shanghai 201306, China}
\affiliation{Department of Physics, Royal Institute of Technology
(KTH), SE-10691 Stockholm, Sweden }\affiliation{Department of
Physics and Shanghai Key Laboratory of Particle Physics and
Cosmology, Shanghai Jiao Tong University, Shanghai 200240, China }

\author{C. Qi \footnote{chongq@kth.se}}
\affiliation{Department of Physics, Royal Institute of Technology
(KTH), SE-10691 Stockholm, Sweden }

\author{Y. Lei}
\affiliation{Key Laboratory of Neutron Physics, Institute of Nuclear
Physics and Chemistry, China Academy of Engineering Physics,
Mianyang 621900, China}

\author{R. Liotta}
\affiliation{Department of Physics, Royal Institute of Technology
(KTH), SE-10691 Stockholm, Sweden }

\author{R. Wyss}
\affiliation{Department of Physics, Royal Institute of Technology
(KTH), SE-10691 Stockholm, Sweden }

\author{Y. M. Zhao \footnote{Corresponding author: ymzhao@sjtu.edu.cn}}
\affiliation{Department of Physics and Shanghai Key Laboratory of
Particle Physics and Cosmology, Shanghai Jiao Tong University,
Shanghai 200240, China } \affiliation{Center of Theoretical Nuclear
Physics, National Laboratory of Heavy Ion Accelerator, Lanzhou
730000, China}

\date{\today}

\begin{abstract}
The low-lying level schemes and electromagnetic transitions of
$^{109}$Te, $^{109}$I, and the neighboring even-even nucleus
$^{108}$Te, are calculated within the framework of the $SD$-pair
approximation of the nuclear shell model.  Good agreement is
obtained between the calculated results and experimental data. The
favored components of low-lying bands are discussed in the
collective nucleon-pair subspace. The weak-coupling picture shown in
these nuclei and its relationship with residual
quadrupole-quadrupole interaction between valence protons and
neutrons are analyzed.

\end{abstract}


\vspace{0.4in}

\maketitle

\newpage

\section{Introduction}

Great experimental and theoretical efforts have been made in recent
years to study the structure and decay properties of
neutron-deficient tin, tellurium, iodine and xenon isotopes near the
$N=Z=50$ closed shells \cite{Review,
Sn-104,Te108,Te109,I109-2,Pro13,San07,SP-Sn101,
Jiang-Sn,Qi12,Sn-core-Excitation-1,Mor11}. Towards the proton drip
line, these nuclei become unstable against particle emissions
\cite{Te105-alpha,I109-proton,Theo-alpha-2,Theo-alpha-3}. $\alpha$
decays have been observed in nuclei $^{105-110}$Te and $^{108-113}$I
\cite{Review, Te105-alpha}, and the nucleus $^{109}$I is known as
a proton emitter \cite{I109-proton,Theo-alpha-3}. Another feature of
particular interest in this mass region is the behavior of the band
structure and electromagnetic transition properties in relation to
the doubly magic nucleus $^{100}$Sn.  The spectroscopic studies
suggest a vibrational-like collective character in even-even
tellurium nuclei \cite{energy,Te106}. The manifestation of vibrational
collectivity in these nuclei is, however, not supported by $B(E2)$
measurements \cite{Mol05}. Octupole correlations were found in the
nuclei $^{108,109}$Te \cite{Te108-8,Te109-octupole}. Bands built on
the $\nu h_{11/2}$ and $\pi h_{11/2}$ orbits are systematically
observed
 in the odd-mass tellurium and iodine isotopes. The
low-lying $\nu h_{11/2}$ bands in odd-mass tellurium isotopes follow
the same trend as those yrast states in the even-even core
\cite{Te109,Te111,Te107,Te113}. This was explained in terms of
core-particle coupling \cite{Te111}. The $\pi h_{11/2}$ bands for
odd-mass iodine isotopes reflect a decrease in quadrupole
deformation moving away from the midshell, with the maximum
occurring in $^{117,119}$I \cite{I109-1}.

The measurement of electromagnetic transitions in this nuclear
region is a challenging task due to the very small reaction
cross-sections leading to the nuclei of interest. So far, the
lightest tin, tellurium and iodine nuclei with known reduced
transition probabilities are $^{104}$Sn \cite{Sn-104}, $^{108}$Te
\cite{Te108}, $^{109}$Te \cite{Te109} and $^{109}$I \cite{I109-2}.
Few theoretical studies have been carried out to analyze the band
structures and electromagnetic transitions in the nuclei $^{109}$Te
and $^{109}$I. Among these works, cranked Woods-Saxon calculations
\cite{I109-1} interpreted the band structures of $^{109}$I as being
built on the $\pi g_{7/2}$ and $\pi h_{11/2}$ states in a weakly
triaxial deformed nucleus. Interacting boson-fermion model
calculations \cite{Te109-IBFM} discussed the band structures of
$^{109}$Te and identified two favored bands built on the $\nu
g_{7/2}$ and $\nu h_{11/2}$ neutron quasiparticle states.
Shell-model calculations with the realistic CD-Bonn nucleon-nucleon
potential on $^{109}$Te \cite{Te109} and $^{109}$I \cite{I109-2}
reproduced the experimental excitation energies, but showed large
deviations from the experimental $B(E2)$ strengths in certain cases.

Recent lifetime measurements on $^{108}$Te \cite{Te108}, $^{109}$Te
\cite{Te109} and $^{109}$I \cite{I109-2} showed that the $B(E2)$
values are approximately equal. This suggests that the additional
proton (or neutron) in $^{109}$I (or $^{109}$Te) might have
negligible effect on the reduced transition probabilities. Based on
shell-model calculations, it was speculated that in these states the
additional unpaired nucleon is weakly coupled to the even-even core
$^{108}$Te \cite{Te109,I109-2}.

The purpose of this paper is to study the low-lying band structures
and electromagnetic transitions in the nuclei $^{108,109}$Te and
$^{109}$I within the framework of the nucleon pair approximation
(NPA). The NPA \cite{chen-npa} has been shown to be a reliable and
economic approximation of the shell model. It has been successfully
applied to describe even-even, odd-$A$ and odd-odd nuclei with
$A\sim 80$ \cite{NPA-cal-80}, $100$ \cite{Jiang-Sn}, $130$
\cite{NPA-cal-130} and $210$ \cite{NPA-cal-210}. In this model, the
dimension of the collective nucleon-pair subspace is small, thus
providing a simple and illuminating picture of the
structure of the nuclei under investigation. In particular it allows one to
evaluate the probability of the existence of weak-coupling schemes
in a straightforward manner.

This paper is organized as follows. In Sec. II we give a brief
introduction to the nucleon pair approximation (NPA), including the
basis, the Hamiltonian, the transition operators, and the
parametrization of our calculations. In Sec. III we  present our
calculations on the excitation energies of the low-lying states,
their dominant configurations and the electromagnetic transition
properties. Our summary and conclusion are given in Sec. IV.

\section{Theoretical Framework}
For medium-heavy nuclei, the dimension of the shell model
configuration space is usually prohibitively large and one must
resort to various truncation schemes, e.g. the interacting boson
model \cite{IBM}, the broken pair approximation \cite{Dpair}, the
fermion dynamical symmetry model \cite{FDSM} as well as the NPA
\cite{chen-npa}. In the NPA approach, a collective pair with angular
momentum $r$ and projection $M$ is defined as \cite{chen-npa}
\begin{eqnarray}
&& A^{(r)\dagger}_{M \sigma} = \sum_{j_{\sigma} j'_{\sigma}}
y(j_{\sigma} j'_{\sigma}r) \left( C^{\dagger}_{j_{\sigma}} \times
C^{\dagger}_{j'_{\sigma}} \right)_M^{(r)} ~,  \nonumber
\end{eqnarray}
where $C^{\dagger}_{j_{\sigma}}$ is the single-particle creation
operator in the $j$ orbit, and $\sigma=\pi$ and $\nu$ is the index
of proton and neutron degrees of freedom, respectively. $r=0, 2, 4,
6, 8$ corresponds to $S$, $D$, $G$, $I$ and $K$ pairs. The numbers
$y(j_{\sigma} j'_{\sigma}r)$ are the so-called structure
coefficients of the nucleon pair with spin $r$.

In an even-even system with $2N$ valence protons or neutrons, we
assume that all the valence nucleons are coupled to collective
pairs. Our collective nuclear pair subspace is constructed by
coupling $N$ collective pairs $r_{1},r_{2}\cdots r_{N}$ stepwise,
\begin{eqnarray}
    & A^{(J_{N})\dagger}_{M_{J_{N}}}(r_{1}r_{2}\cdots r_{N},J_{1}J_{2}\cdots J_{N})|0\rangle \equiv A^{(J_{N})\dagger}_{M_{J_{N}}}|0\rangle\nonumber\\
     ~~~&=( \cdots ( A^{r_1 \dagger }
    \times A^{r_2 \dagger} )^{(J_2)} \times \cdots \times A^{r_N \dagger})^{(J_N)}_{M_{J_{N}}}|0\rangle  ~.\nonumber
\end{eqnarray}
Similarly, for an odd-$A$ system with $2N+1$ valence protons or
neutrons, all valence nucleons are paired except the last nucleon, which can
occupy any single-particle level $j$ of the shell model space
under consideration. Our nucleon-pair subspace is given by
successively coupling the $N$ nucleon pairs to the unpaired nucleon
in a single-$j$ orbit as
\begin{eqnarray}
     &A^{(J_{N})\dagger}_{M_{J_{N}}}(jr_{1}r_{2}\cdots r_{N},J_{1}J_{2}\cdots J_{N})|0\rangle \equiv A^{(J_{N})\dagger}_{M_{J_{N}}}|0\rangle\nonumber\\
     &=( \cdots ((C_{j}^{\dagger}\times A^{r_1 \dagger })^{(J_{1})}
    \times A^{r_2 \dagger} )^{(J_2)} \times \cdots \times A^{r_N \dagger})^{(J_N)}_{M_{J_{N}}}|0\rangle  ~,\nonumber
\end{eqnarray}
where
$J_{i}$ (half integer) denotes the total angular momentum for the
first $2i+1$ nucleons.

As in Ref. \cite{chen-npa}, we choose  in our calculations a
complete set of non-orthonormal but linearly independent many-pair
basis states. If the basis states are chosen appropriately, all
the eigenvalues of the overlap matrix $\langle
0|A^{(J'_{N})}_{M_{J'_{N}}}A^{(J_{N})\dagger}_{M_{J_{N}}}|0\rangle$
are non-zero. This practice guarantees that all multi-pair basis states in the NPA
calculations are not over-complete.

\begin{table}
\caption{ Single-particle  (s.p.)  energies $\epsilon_{j_{\sigma}}$
(in MeV) and two-body interaction parameters $G^0_{\sigma} $,
$G^2_{\sigma} $, $\kappa_{\sigma} $, $\kappa_{\pi \nu} $. The unit
of $G^0_{\sigma} $ is MeV; the units of $G^2_{\sigma} $,
$\kappa_{\sigma} $ and $\kappa_{\pi \nu} $ are MeV/$r_{0}^{4}$,
$r_0^2=1.012A^{1/3}$ fm$^{2}$. $\sigma=\pi, \nu$ stands for proton
and neutron, respectively. }
\begin{tabular}{cccccccccccccccccccccccccccccccccccccccc}
 \hline
 \hline
 $j $ &$~$$s_{1/2}$&$~$$d_{3/2}$&$~$$d_{5/2}$&$~$$g_{7/2}$&$~$$h_{11/2}$\\
 $\epsilon_{j_{\pi}}$ &1.550&1.660&0.172&0.000&3.550\\
 $\epsilon_{j_{\nu}}$ &1.550&1.660&0.172&0.000&3.550\\
\hline\\
$G^0_{\nu}$ &$~$$G^2_{\nu}$&$~$${\kappa}_{\nu} $&$~$$G^0_{\pi}$&$~$$G^2_{\pi} $&$~$${\kappa}_{\pi} $&$~$${\kappa}_{\pi \nu} $\\
$-0.18$           &$-0.036$           &$-0.015$ &$-0.20$  &$-0.036$ &$-0.0125$ &$-0.05$         \\
 \hline \hline
\end{tabular}
\label{table:1}
\end{table}

\begin{figure*}
\includegraphics[width = 0.92\textwidth]{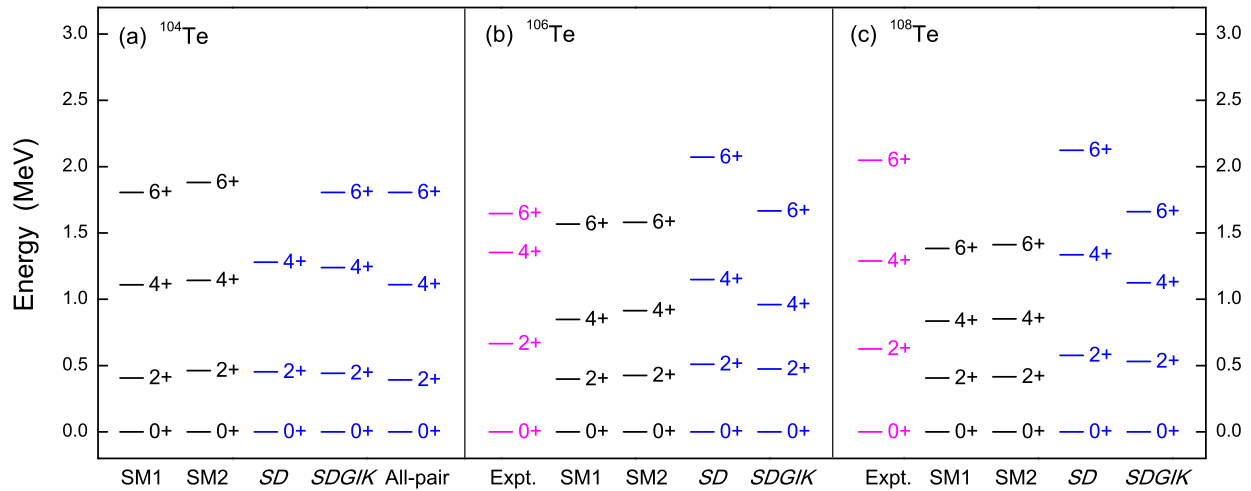}
\caption{(Color online) Ground state bands in (a) $^{104}$Te, (b)
$^{106}$Te and (c) $^{108}$Te. The experimental data are taken from
Ref. \cite{energy}. The shell-model calculations without isospin
symmetry (denoted by SM1), the $SD$-pair approximation (denoted by
$SD$), the $SDGIK$-pair approximation (denoted by $SDGIK$) and the
calculations in the subspace constructed by all possible NPA pairs
(denoted by All-pair) are performed with the same phenomenological
interactions in Eq (\ref{eq1}). The shell model calculations with
isospin symmetry (denoted by SM2) are shown for comparison.
\label{figSM} }
\end{figure*}

The NPA Hamiltonian is chosen to have the form
\begin{eqnarray}
H & = & \sum_{j_{\sigma}} \epsilon_{j_{\sigma} }C^{\dagger}_{j_{\sigma}}  C_{j_{\sigma}} \nonumber\\
&+& \sum_{\sigma} \left(G^0_{\sigma} {\cal
P}^{(0)\dagger}_{\sigma}\cdot {\cal
P}^{(0)}_{\sigma}+G^2_{\sigma}{\cal P}^{(2) \dagger }_{\sigma} \cdot {\cal P}^{(2)}_{\sigma} \right) \nonumber\\
& + & \sum_{\sigma}  \kappa_{\sigma} Q_{\sigma}\cdot Q_{\sigma}+
\kappa_{\pi\nu}Q_{\pi}\cdot Q_{\nu},\label{eq1}
\end{eqnarray}
where $\epsilon_{j_{\sigma}}$ is the single-particle energy,
$G^0_{\sigma}$, $G^2_{\sigma}$, $\kappa_{\sigma}$ and
$\kappa_{\pi\nu}$ are the two-body interaction strengths
corresponding to monopole, quadrupole pairing  and
quadrupole-quadrupole interactions between valence nucleons. We have
\begin{eqnarray}
&& {\cal P}^{(0)\dagger}_{\sigma} =   \sum_{j_{\sigma}}
\frac{\sqrt{2{j}_{\sigma}+1}}{2}(C_{j_{\sigma}}^{\dagger} \times
C_{j_{\sigma}}^{\dagger})^{(0)}_0 ,\nonumber \\
 && {}{\cal P}^{(2)\dagger}_{\sigma } = \sum_{j_{\sigma} j'_{\sigma}}
q(j_{\sigma} j'_{\sigma}) \left( C^{\dagger}_{j_{\sigma}} \times
C^{\dagger}_{j'_{\sigma}} \right)_M^{(2)},\nonumber \\
&& {}Q_{\sigma } = \sum_{j_{\sigma }j'_{\sigma }} q(j_{\sigma }
j'_{\sigma }) \left( C^{\dagger}_{j_{\sigma }} \times
\tilde{C}_{j'_{\sigma }} \right)^{(2)}_M ,  \nonumber
\end{eqnarray}
where $q(jj') = \frac{(-)^{j-1/2}}{\sqrt{20\pi} }\hat{ j} \hat{j'}
C_{j1/2, j' -1/2}^{20}\langle n l|r^2 |n l' \rangle$.  $ C_{j 1/2,j'
-1/2}^{2 0} $ is the Clebsch-Gordan coefficient. The isospin
symmetry is not conserved in our NPA Hamiltonian. But it has to be
pointed out that the effect of isospin mixture on the low-lying
states is small for the nuclei treated here. To illustrate this
point, we make comparison of ground state bands for
$^{104,106,108}$Te between the shell model calculations without
(denoted by SM1) and with isospin symmetry (denoted by SM2) in Fig.
\ref{figSM}. It is seen that both calculations give rather similar
results.

The single-particle energies and two-body interaction parameters in
our calculations are shown in Table \ref{table:1}. The neutron
single-particle energies of $g_{7/2}$ and $d_{5/2}$ orbitals are
taken from the experimental excitation energies in $^{101}$Sn
\cite{SP-Sn101}. There are no experimental data for the remaining
orbitals. Their single-particle energies are extracted from a shell
model calculation \cite{SM-Sn}. The proton single-particle energies
are taken to be the same as those for neutrons.

There are totally seven parameters for the two-body interactions:
$G^0_{\pi} $, $G^0_{\nu} $, $G^2_{\pi} $, $G^2_{\nu} $,
$\kappa_{\pi}$, $\kappa_{\nu} $ and $\kappa_{\pi \nu} $. For
$^{109}$Te and $^{109}$I, we assume the same parameters as their
even-even core $^{108}$Te. We take $G^0_{\nu}=-0.18 $ MeV, which is
the same value as the one used in Ref. \cite{Jiang-Sn}. As the
proton number is close to the neutron number in this region we
adopt for the strength of the proton interaction the value
$G^0_{\pi}$ = $-0.20$ MeV.  The remaining five parameters are
obtained by fitting to the excitation energies and $B(E2)$ values in
nuclei $^{108,109}$Te and $^{109}$I.

\begin{figure*}
\includegraphics[width = 0.77\textwidth]{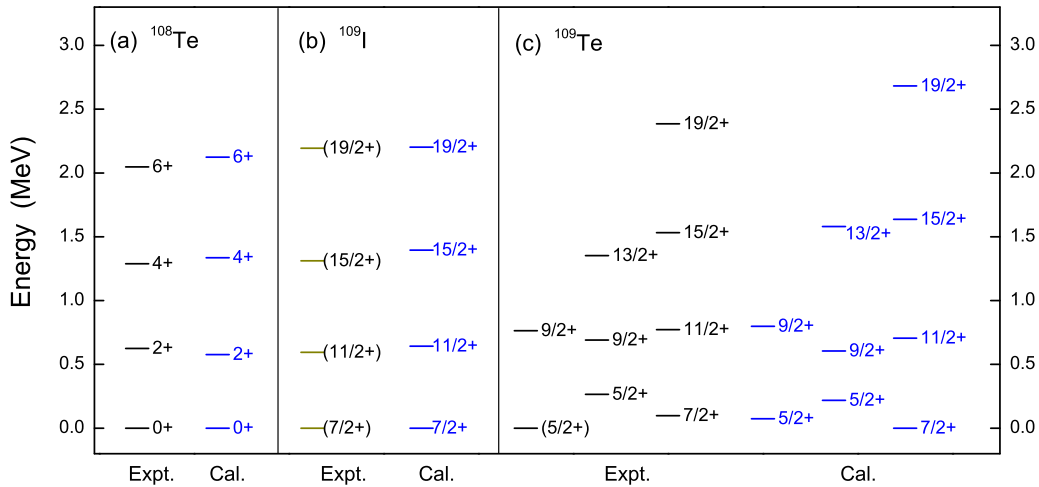}
\caption{(Color online) Partial level schemes for low-lying
positive-parity states in (a) $^{108}$Te, (b) $^{109}$I and (c)
$^{109}$Te. The experimental data of $^{108,109}$Te and $^{109}$I
are taken from Refs. \cite{energy} and \cite{I109-1}, respectively.
\label{fig1} }
\end{figure*}

\begin{figure*}
\includegraphics[width = 0.63\textwidth]{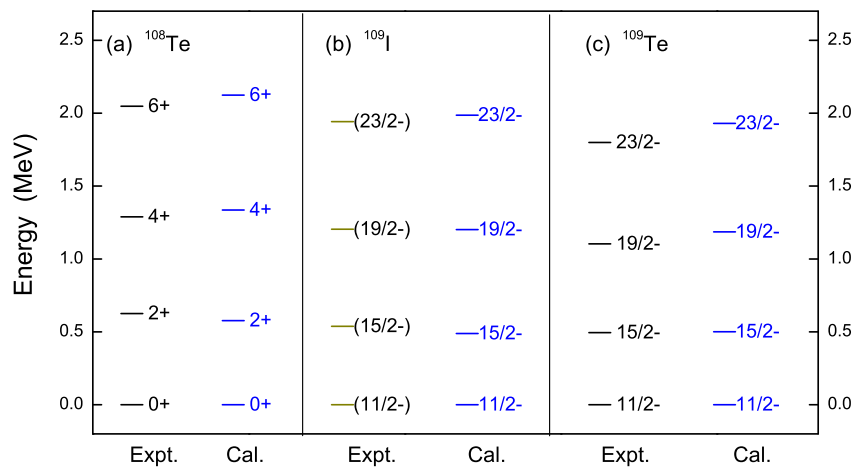}
\caption{(Color online) Same as Fig. \ref{fig1} but for low-lying
$h_{11/2}$ bands of $^{109}$I and $^{109}$Te in comparison with the
ground state band in $^{108}$Te, plotted relative to the bandheads.
\label{fig2} }
\end{figure*}

The $E2$ transition operator is defined  by $  T(E2) = e_{\pi}
Q_{\pi} + e_{\nu}Q_{\nu}, $ where $e_{\pi}$ and $e_{\nu}$ are the
effective charges of valence protons and neutrons, respectively. The
$B(E2)$ value in unit of $\rm e^2 fm^4$ is given by
\begin{eqnarray}
B(E2,J_{i}\rightarrow J_{f})=\frac{2J_{f}+1}{2J_{i}+1}(e_{\pi}
\chi_{\pi}+e_{\nu} \chi_{\nu})^2 r_0^{4}~~,\label{eq2}
\end{eqnarray}
with reduced matrix element $\chi_{\sigma}=\langle
\beta_{f},J_{f}||Q_{\sigma}||\beta_{i},J_{i}\rangle$ ($\sigma=\pi,
\nu$) and $r_0^2=1.012A^{1/3}$ fm$^{2}$. $|\beta_{i},J_{i}\rangle$
is the eigenfunction of $J_{i}$ state. Our neutron effective charge
is taken to be $e_{\nu}=1.28 e$, the same as for tin isotopes
\cite{Jiang-Sn}. The proton effective charge $e_{\pi}=1.79 e$ is
obtained by fitting to experimental data.

The $M1$ transition operator is defined by $ T(M1) = g_{l\pi}
l_{\pi} + g_{l\nu} l_{\nu} + g_{s\pi} s_{\pi} + g_{s\nu} s_{\nu}$,
where $l_{\sigma}$ and $s_{\sigma}$ are the orbital and spin angular
momenta, $g_{l \sigma}$ and $g_{s \sigma}$ are the effective orbital
and spin gyromagnetic ratios, respectively. The effective spin
gyromagnetic ratios are taken to be $g_{s\pi}=5.586\times0.7 ~
\mu_N$ and $g_{s\nu}=-3.826\times0.7 ~ \mu_N$, where the number
$0.7$ is the conventional quenching factor (see also Ref.
\cite{Qi12}). Two sets of effective orbital gyromagnetic ratios are
used in this paper. In the first set, we use their free values, i.e.
$g_{l\nu}=0 ~ \mu_N$ and $g_{l\pi}=1 ~ \mu_N$. In the other set, we
take $g_{l\nu}=0 ~ \mu_N$ and $g_{l\pi}=1.35 ~ \mu_N$, which are the
optimized parameters determined by fitting to the experimental data
in this region.

Our nucleon pair subspace is constructed by $SD$  pairs of valence
protons and neutrons, with respect to the doubly-closed shell
nucleus $^{100}$Sn. We have also investigated the $SDGIK$-pair
subspace and found that the $G, I, K$ pairs do not contribute
significantly to the low-lying states of the nuclei $^{108,109}$Te
and $^{109}$I. As in Ref. \cite{NPA-cal-130}, we use the BCS pairs
as our $S$ pair. The $D$ pair is obtained by using the commutator
$D^{\dag}=\frac{1}{2}[Q ,S^{\dag}]$ \cite{Dpair}.

\begin{figure}
\includegraphics[width = 0.481\textwidth]{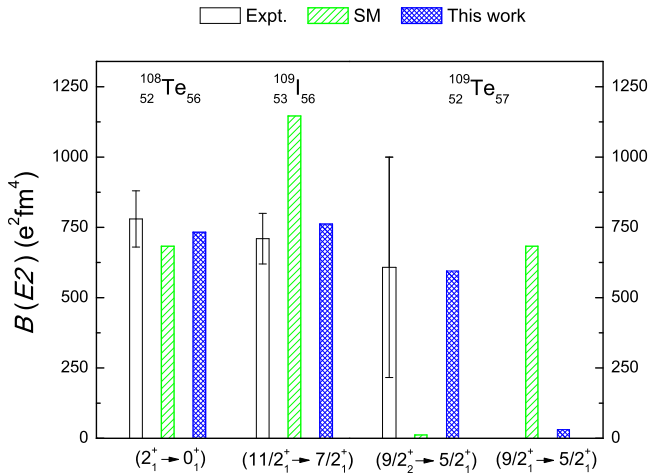}
\caption{(Color online) Comparison between theoretical $B(E2)$
transitions and the experimental data for nuclei $^{108}$Te
\cite{Te108}, $^{109}$I \cite{I109-2} and $^{109}$Te \cite{Te109}.
The shell model results (SM) are taken from Refs.
\cite{I109-2,Te109}. One sees the experimental ground-state
transitions between $^{108}$Te, $^{109}$I and $^{109}$Te are
approximately equal, which indicates that the extra proton (or
neutron) in $^{109}$I (or $^{109}$Te) has no significant effect on
the reduced transition probabilities for these states. \label{fig3}
}
\end{figure}

It was shown in Ref. \cite{validity} that the NPA is an efficient
truncation scheme of the shell model for nuclei $^{44,46}$Ca,
$^{130,131}$Te and $^{132}$I by using phenomenological as well as
realistic interactions. In order to probe the validity of the NPA
even in the nuclei to be studied here we also perform the shell
model as well as the NPA calculations and compared the results. In
Fig. \ref{figSM}, we make comparisons of ground state bands for
$^{104}$Te, $^{106}$Te and $^{108}$Te calculated by $SD$-pair
approximation (denoted by $SD$), $SDGIK$-pair approximation (denoted
by $SDGIK$), the calculation in the subspace constructed by all
possible NPA pairs (denoted by All-pair) and the shell model
(denoted by SM1) by taking the same phenomenological interactions in
Eq (\ref{eq1}). It is shown that the results of the $SD$ reasonably
agree with those of the SM1 (especially for the $2_1^+$ state),
indicating that the $SD$ pairs are very important in building up
low-lying states. For the nucleus $^{108}$Te, our $SD$ NPA
calculation noticeably overestimates the excitation energy of the
yrast $6^+$ state in comparison with that of the shell model, due to the influence of other pairs,
namely $G, I, K$. If all possible NPA pairs are taken into account
(see Fig. \ref{figSM}(a) for $^{104}$Te), our results are equivalent
to those of the SM1. Moreover, it is to be pointed out that the
remarkable feature in Figs. \ref{fig1} and \ref{fig2} is that the
$SD$-pair approximation agrees well with the experiment. This
indicates that the $SD$-pair approximation with the phenomenological
interactions is very well fitted to explain low-lying states in
these nearly spherical nuclei, especially the energies and $E2$
transition properties of the $2^+_1$ state of concern (see Fig.
\ref{fig3}, Tables \ref{table:2} and \ref{table:4}). A much more
sophisticated treatment of the effective interaction may be
necessary for calculations in a larger space with more pairs or with
in the shell-model framework in order to get a better agreement with
experimental data.

\section{Calculations and discussions}

\begin{table}
\caption{The $B(E2,J_{i}\rightarrow J_{f})$ values in units of
$e^{2} \rm fm^{4}$. The columns $\chi_{\sigma}$ ($\sigma=\pi, \nu$)
show the reduced matrix elements of Eq. (\ref{eq2}). The
experimental data and shell model results are taken from $^{\rm{a}}$
Ref. \cite{Te108}, $^{\rm{b}}$ Ref. \cite{I109-2} and $^{\rm{c}}$
Ref. \cite{Te109}.}
\begin{tabular}{c cc c cc c cc c cc c cccccccccccccc}
\hline \hline
  &$J_{i}$  &$J_{f}$                  &Expt. &NPA &SM &$\chi_{\pi}$ &$\chi_{\nu}$\\
\hline\\
$^{108}$Te &&\\
&$2^+_1$ &$0^+_1$&$780(^{+100}_{-80})^{\rm{a}}$&$733$&$680^{\rm{c}}$&$3.429$&$5.009$  \\
&$4^+_1$ &$2^+_1$ & $-$ &$894$&$-$&$1.633$&$4.215$  \\

$^{109}$I&&\\
  &$11/2^+_1$ &$7/2^+_1$ &$710(90)^{\rm{b}}$    &$762$&$1146^{\rm{b}}$   &$1.757$&$3.003$ \\
&$15/2^+_1$ &$11/2^+_1$ &$-$    &$1034$ &$-$  &$1.625$&$3.724$ \\
&$15/2^-_1$ &$11/2^-_1$ &$-$    &$880$ &$-$  &$1.986$&$2.753$ \\
&$19/2^-_1$ &$15/2^-_1$ &$-$    &$983$ &$-$  &$1.473 $&$3.603$ \\

$^{109}$Te&&\\
  &$9/2^+_2$ &$5/2^+_1$ &$608(392)^{\rm{c}}$    &$595$ &$12^{\rm{c}}$  &$2.096$&$2.153$ \\
  &$9/2^+_1$ &$5/2^+_1$ &$-$    &$31$ &$683^{\rm{c}}$  &$0.443$&$0.550$ \\
  &$9/2^+_1$ &$5/2^+_2$ &$340(93)^{\rm{c}}$    &$56$ &$6^{\rm{c}}$  &$0.585$&$0.737$ \\
    &$9/2^+_2$ &$5/2^+_2$ &$-$    &$0.46$ &$407^{\rm{c}}$  &$-0.063$&$0.231$ \\
&$13/2^+_1$ &$9/2^+_1$ &$-$    &$310$ &$850^{\rm{c}}$  &$0.785$&$2.270$ \\
&$11/2^+_1$ &$7/2^+_1$ &$-$    &$583$ &$810^{\rm{c}}$  &$1.804$&$2.254$ \\
&$15/2^+_1$ &$11/2^+_1$ &$-$    &$719$ &$-$  &$1.420$&$3.016$ \\
&$15/2^-_1$ &$11/2^-_1$ &$-$    &$861$ &$-$  &$1.810$&$2.941$ \\
&$19/2^-_1$ &$15/2^-_1$ &$-$    &$1005$ &$-$  &$1.458$&$3.686$ \\

\hline \hline\\
\end{tabular}
\label{table:2}
\end{table}

\begin{table}
\caption{The $B(M1,J_{i}\rightarrow J_{f})$ values of $^{109}$Te in
units of $10^{-3} \rm \mu_N^{2}$. For the proton effective orbital
gyromagnetic ratio, we adopt the value $g_{l\pi}=1 ~ \mu_N$ in
``NPA-1'' and $g_{l\pi}=1.35 ~ \mu_N$ in ``NPA-2''. The experimental
data and shell model results are taken from Ref. \cite{Te109}.}
\begin{tabular}{c cc c cc c cc c cc c cccccccccccccc}
\hline \hline
  &$J_{i}$  &$J_{f}$                  &Expt. &NPA-1 &NPA-2 &SM \\
\hline

  &$7/2^+_1$ &$5/2^+_1$ &$-$    &$3$&$2$ &$151$  &$$&$$ \\
  &$7/2^+_1$ &$5/2^+_2$ &$-$    &$6$&$5$ &$55$  &$$&$$ \\
  &$7/2^+_2$ &$5/2^+_1$ &$-$    &$4$&$10$ &$2$  &$$&$$ \\
&$7/2^+_2$ &$5/2^+_2$ &$-$    &$0.2$&$13$ &$76$  &$$&$$ \\
&$9/2^+_1$ &$7/2^+_1$ &$6(1)$    &$1$&$10$ &$162$  &$$&$$ \\
&$9/2^+_1$ &$7/2^+_2$ &$-$    &$24$&$150$ &$126$  &$$&$$ \\
&$9/2^+_2$ &$7/2^+_1$ &$-$    &$16$&$15$ &$0$  &$$&$$ \\
&$9/2^+_2$ &$7/2^+_2$ &$137(92)$    &$57$&$127$ &$102$  &$$&$$ \\

\hline \hline\\
\end{tabular}
\label{table:4}
\end{table}

\begin{table}
\caption{ Same as Table \ref{table:4} except for magnetic moments
$\mu$ (in units of $\mu_N$) predicted in this work. }
\begin{tabular}{ccccc cccccc}\hline \hline
\multicolumn{1}{c}{}    &\multicolumn{2}{c}{NPA-1}
&\multicolumn{2}{c}{NPA-2 } &\multicolumn{2}{c}{}
&\multicolumn{2}{c}{NPA-1}  &\multicolumn{2}{c}{NPA-2}\\

\hline
$^{108}$Te   &           &               &               &               &&$^{109}$I    &           &               &           &       \\
$2_1^+$     &+0.910    &                &+1.237 &&&$7/2_1^+$&+2.282&&+3.592&\\
$4_1^+$     &+1.001    &                &+1.331 &&&$11/2_1^+$     &+2.954  &     &+4.510&\\
\\

$^{109}$Te   &           &               &               &               &&    &           &               &           &       \\
$5/2_1^+$     &$-1.096$    &&$-1.078$&&&$5/2_2^+$&+0.568&&+0.593&\\
$9/2_1^+$     &+1.582    &&+1.888    &&&$9/2_2^+$     &+0.269  &     &+0.737&\\
$7/2_1^+$     &+1.034    &&+1.051    &&&$11/2_1^+$     &+2.008&&+2.395&\\

\hline \hline\\
\end{tabular}
\label{table:mu}
\end{table}

\begin{table*}
\caption{Absolute values of overlaps between the calculated NPA
low-lying states of $^{109}$I (or $^{109}$Te) and the corresponding
weak-coupling wave function $|({\sigma}j) \otimes \rm ^{108}Te
(\it{\lambda});k \rangle$.  $k$ and $\lambda$ correspond to the
state of odd-mass nucleus and its neighboring even-even core
$^{108}$Te, respectively. ${\sigma}j$ ($\sigma=\pi, \nu$) refers to
the unpaired valence nucleon in a single-$j$ orbit.
 }
\begin{tabular}{ccccccccccccccccccccccc}
 \hline
 \hline\\
 &&&$^{109}$I&&&&&&&&$^{109}$Te\\
\cline{2-6} \cline{9-15}\\
     &$|\rm ^{109}I (\it{k})\rangle$        &$~$     & $|({\pi}j)\otimes \rm ^{108}Te (\it{\lambda});k\rangle$         &$~$     &Overlap  &&& &$|\rm ^{109}Te (\it{k})\rangle$       &$~$     & $|({\nu}j)\otimes \rm ^{108}Te (\it{\lambda});k\rangle$         &$~$     &Overlap &$~$           \\

 %\hline

&$|7/2_{1}^{+}\rangle$        && $|({\pi}g_{7/2})\otimes(0^{+}_{1})\rangle$&&$0.82$        &&&& $|5/2_{1}^{+}\rangle$ & & $|({\nu}d_{5/2})\otimes(0^{+}_{1})\rangle$ &&$0.94$  &\\
&$|11/2_{1}^{+}\rangle$       && $|({\pi}g_{7/2})\otimes(2^{+}_{1})\rangle$&&$0.86$        &&&&$|9/2_{2}^{+}\rangle$  & & $|({\nu}d_{5/2})\otimes(2^{+}_{1})\rangle$ &&$0.89$ &\\
\cline{9-15}
&$|15/2_{1}^{+}\rangle$       && $|({\pi}g_{7/2})\otimes(4^{+}_{1})\rangle$&&$0.87$        &&&&$|5/2_{2}^{+}\rangle$  & & $|({\nu}g_{7/2})\otimes(2^{+}_{1})\rangle$ &&$0.81$ &\\
&$|19/2_{1}^{+}\rangle$       && $|({\pi}g_{7/2})\otimes(6^{+}_{1})\rangle$&&$0.93$        &&&&$|9/2_{1}^{+}\rangle$   & & $|({\nu}g_{7/2})\otimes(2^{+}_{1})\rangle$ &&$0.89$&\\
\cline{2-6}
&$|11/2_{1}^{-}\rangle$       &&$|({\pi}h_{11/2})\otimes(0^{+}_{1})\rangle$&&$0.85$        &&&&$|13/2_{1}^{+}\rangle$ & & $|({\nu}g_{7/2})\otimes(4^{+}_{1})\rangle$&&$0.77$&\\
\cline{9-15}
&$|15/2_{1}^{-}\rangle$       &&$|({\pi}h_{11/2})\otimes(2^{+}_{1})\rangle$&&$0.91$        &&&&$|7/2_{1}^{+}\rangle$ & & $|({\nu}g_{7/2})\otimes(0^{+}_{1})\rangle$ &&$0.93$&\\
&$|19/2_{1}^{-}\rangle$       &&$|({\pi}h_{11/2})\otimes(4^{+}_{1})\rangle$&&$0.92$        &&&&$|11/2_{1}^{+}\rangle$ & & $|({\nu}g_{7/2})\otimes(2^{+}_{1})\rangle$ &&$0.92$&\\
&$|23/2_{1}^{-}\rangle$       &&$|({\pi}h_{11/2})\otimes(6^{+}_{1})\rangle$&&$0.95$        &&&&$|15/2_{1}^{+}\rangle$ & & $|({\nu}g_{7/2})\otimes(4^{+}_{1})\rangle$ &&$0.89$&\\
&&&&&                                                                                  &&&&$|19/2_{1}^{+}\rangle$ & & $|({\nu}g_{7/2})\otimes(6^{+}_{1})\rangle$ &&$0.93$&\\

\cline{9-15}
&&&&&                                                                                  &&&&$|11/2_{1}^{-}\rangle$ & & $|({\nu}h_{11/2})\otimes(0^{+}_{1})\rangle$ &&$0.95$&\\
&&&&&                                                                                  &&&&$|15/2_{1}^{-}\rangle$ & & $|({\nu}h_{11/2})\otimes(2^{+}_{1})\rangle$ &&$0.98$&\\
&&&&&                                                                                  &&&&$|19/2_{1}^{-}\rangle$ & & $|({\nu}h_{11/2})\otimes(4^{+}_{1})\rangle$ &&$0.99$&\\
&&&&&                                                                                  &&&&$|23/2_{1}^{-}\rangle$ & & $|({\nu}h_{11/2})\otimes(6^{+}_{1})\rangle$  &&$0.99$&\\

\hline \hline
\end{tabular}
\label{table:3}
\end{table*}

Our calculated low-lying energy levels and electromagnetic
transitions are presented in Figs. \ref{fig1} $\sim$ \ref{fig3} and
Tables \ref{table:2}$-$\ref{table:mu}. The experimental excitation
energies of nuclei $^{108,109}$Te and $^{109}$I are taken from Refs.
\cite{energy,I109-1}. The experimental electromagnetic transitions
are taken from $^{108}$Te \cite{Te108}, $^{109}$Te \cite{Te109} and
$^{109}$I \cite{I109-2}. The experimental
$B(E2,9/2_{1}^{+}\rightarrow 5/2_{1}^{+})$ measurement in $^{109}$Te
is not available at present. We present their theoretical values in
Fig. \ref{fig3} for comparison. The shell model results (SM) are
taken from Refs. \cite{Te109,I109-2}.

One sees in Figs. \ref{fig1} and \ref{fig2} that our calculated
energies reproduce reasonably well the corresponding experimental
values. The relative level schemes of some states in $^{108}$Te and
$^{109}$I are close to each other, suggesting that the additional
proton in $^{109}$I might be weakly coupled to the
even-even core $^{108}$Te \cite{I109-2}. That is, an unpaired proton in a
single-$j$ orbit ($\pi j$) might be coupled to the $\lambda$ state
of $^{108}$Te ($|^{108}$Te$(\lambda)\rangle$) to induce the $k$
state of $^{109}$I, i.e. $|^{109}$I$(k)\rangle=|({\pi}j)\otimes \rm
^{108}$Te$(\lambda);k\rangle$. The nucleus $^{109}$Te shows a similar pattern.

To understand the structures of these states we analyze them within
our collective nucleon-pair subspace. As discussed above, the
resulting NPA wave function for the $k$ state, i.e. $|\rm
^{109}I(\it{k})\rangle$ and $|\rm ^{109}Te(\it{k})\rangle$, contains
many components consisting of $S$ and $D$ pairs plus the unpaired
nucleon which may occupy any single-$j$ orbit in Table
\ref{table:1}. To evaluate the probability that a weakly coupled
state is included in the NPA wave function, we evaluate the
overlaps $\langle \rm ^{109}I (\it{k})|\it ({\pi}j)\otimes \rm
^{108}Te(\it{\lambda});k \rangle$ and $\langle \rm ^{109}Te
(\it{k})|\it ({\nu}j)\otimes \rm ^{108}Te(\it{\lambda});k \rangle$.
The results are given in Table  \ref{table:3}. One sees therein that
for the states of interest, i.e. those in Figs. \ref{fig1} and
\ref{fig2}, the overlaps are indeed very large. This suggests that
these states can be well represented by the weak coupling between a
collective state in the even-even ``core" and the unpaired nucleon
in a single-$j$ orbit. This is also a strong indication of the
vibrational-like character of nuclei in this region, where the
ground as well as low-lying excited states behave like boson degrees
of freedom, practically unperturbed by the presence of the odd
nucleon.

The states of $^{109}$I shown in Figs. \ref{fig1}(b) and
\ref{fig2}(b) are clearly seen to arise from the coupling of the
vibrational-like ground state band in $^{108}$Te with the unpaired
proton in the $\pi g_{7/2}$ (Fig. \ref{fig1}(b)) and $\pi h_{11/2}$
(Fig. \ref{fig2}(b)) orbit. This is indeed confirmed by the large
wave function overlap between the corresponding states in Table
\ref{table:3}.

The analysis of the positive parity bands of $^{109}$Te in Fig.
\ref{fig1}(c) is more involved. In order to understand the relation
between these states and their eventual (if any)  weak coupling
description, one needs to rely on the calculated overlaps of Table
\ref{table:3}. One sees that the states $5/2_1^{+}$ and
$9/2_2^{+}$, forming the first band in Fig. \ref{fig1}(c), are built
upon the coupling of the $\nu d_{5/2}$ orbit with the collective
core states. Instead, the states $5/2_2^{+}$, $9/2_1^{+}$ and
$13/2_1^{+}$, which form the second band in that Figure, are
atypical,  because it arises from the coupling of the $\nu g_{7/2}$
orbit with only the $2^+_1$ and $4^+_1$ states in $^{108}$Te. The
third band is again a weak coupling band, because it arises
as the coupling of the orbit $\nu g_{7/2}$  with all the states in
$^{108}$Te. The negative parity band in Fig. \ref{fig2}(c) is also a
typical weak coupling band, arising from the coupling of the $\nu
h_{11/2}$ orbit with the core states. It is to be pointed out that
our  results largely agree with what was concluded in Refs.
\cite{I109-1,Te109-IBFM,Te109-band}.

Electromagnetic transition is another sensitive probe of the
calculated wave functions. Unfortunately the experimental
observations in this mass region are still scarce and the
corresponding errors are relatively large.  This can be seen in
Fig. \ref{fig3} and Tables \ref{table:2}$-$\ref{table:4}, where the
available experimental $B(E2)$ and $B(M1)$ values as well as a shell
model and our own NPA calculations are shown. In Table \ref{table:2}
we also list our reduced matrix elements $\chi_{\sigma}$
($\sigma=\pi, \nu$) (see Eq. (\ref{eq2})). Our predicted magnetic
dipole moments ($\mu$) of some low-lying states are presented in
Table \ref{table:mu}.

One sees in Fig. \ref{fig3} and Tables \ref{table:2}$-$\ref{table:4}
that our results agree quite well with available experimental data
except for the $B(E2,9/2^{+}_{1}\rightarrow 5/2^{+}_2)$ value in
$^{109}$Te. To explore this further, one might need new experimental
data in addition to the available ones at present.

As seen in Fig. \ref{fig3}, the experimental $B(E2)$ transitions in the nuclei
$^{108}$Te, $^{109}$I and $^{109}$Te are approximately the same.
This is consistent with our previous results on the structure of
these states, because it indicates that the additional nucleon of
$^{109}$I (or $^{109}$Te) is indeed weakly coupled to the even-even core for
these states, as suggested in Refs. \cite{Te109,I109-2}.

One also sees in Fig. \ref{fig3} that the shell model calculations
reproduce well the $B(E2,2_{1}^{+}\rightarrow 0_{1}^{+})$ value in
$^{108}$Te, but overestimate the $B(E2,11/2_{1}^{+}\rightarrow
7/2_{1}^{+})$ value in $^{109}$I, and fail to describe the
$B(E2,9/2_{2}^{+}\rightarrow 5/2_{1}^{+})$ value in $^{109}$Te.
Moreover, in Table \ref{table:2} one sees that the theoretical
$B(E2,9/2_{2}^{+}\rightarrow 5/2_{1}^{+})$ value of the SM, i.e. 12
$\rm{e^{2}fm^{4}}$, is much smaller than the corresponding
experimental data, i.e. 608(392) $\rm{e^{2}fm^{4}}$. As this
experimental value practically coincides with the SM one for the
transition $(9/2_{1}^{+})\rightarrow (5/2_{1}^{+})$ (683
$\rm{e^{2}fm^{4}}$), in Ref. \cite{Te109} it was suggested
that the ordering of the first two calculated excited $9/2^{+}$
states in $^{109}$Te are inverted. This is not the case in our
present calculation.

\begin{figure}
\includegraphics[width = 0.45\textwidth]{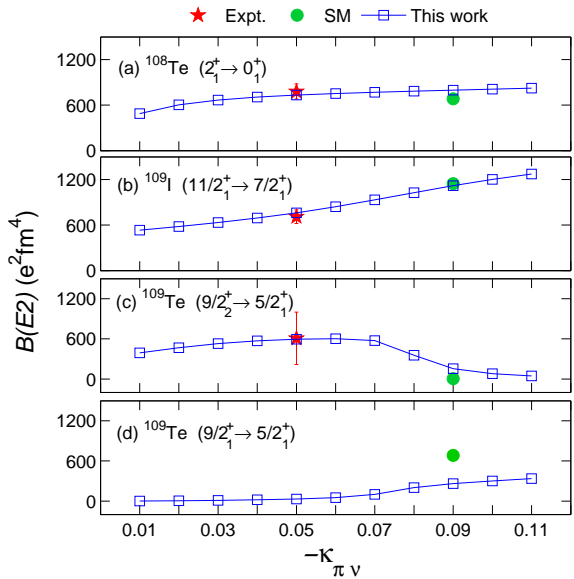}
\caption{(Color online) Same as Fig. \ref{fig3} but for the
theoretical $B(E2)$ values as a function of the
quadrupole-quadrupole proton-neutron interaction strength
$\kappa_{\pi \nu}$. \label{fig4} }
\end{figure}

It thus seems that, as pointed out in Ref. \cite{Te109}, the
presence of the single decoupled valence proton affects the total
measured $B(E2)$ strengths in a manner that is not currently well
understood. To investigate this point further we perform  several
attempts, particularly to survey the sensitivity of the $B(E2)$
values upon the different terms of the interaction entering the
theory. This is not a trivial task, because at the same time we
require  that all the other calculated physical quantities, which
agreed well with available experimental data, should remain
practically unchanged. We could do this very lengthy task because in
our NPA truncated space the computing time needed to perform the
calculations is relatively short. In this search we finally find
that those requirements are fulfilled if one varies the residual
proton-neutron quadrupole-quadrupole interaction ($\kappa_{\pi\nu}$)
in a range from $-0.01$ to $-0.11$ MeV/$r_{0}^{4}$. Our results are
shown in Fig. \ref{fig4}. There are two remarkable features in this
Figure. First, our calculation reproduces the available experimental
data in all the nuclei analyzed here, i.e. $^{108}$Te, $^{109}$Te
and $^{109}$I, by using a strength $\kappa_{\pi \nu}\sim -0.05$
MeV/$r_{0}^{4}$. Instead, the shell model results
\cite{Te109,I109-2}, which agree with experiment only in the nucleus
$^{108}$Te, are reproduced by using $\kappa_{\pi \nu}\sim -0.09$
MeV/$r_{0}^{4}$. This suggests that, as speculated in Ref.
\cite{Te109,I109-2}, the quadrupole-quadrupole correlation in the
realistic shell model interaction might be too strong for nuclei in
this region. It also indicates that $\kappa_{\pi \nu}$ for these
three nuclei is not as strong as the values predicted by empirical
formulas (see Appendix B in Ref. \cite{zhao-npa-cal}), i.e.
$\kappa_{\pi \nu} = -0.08\sim-0.10$ MeV/$r_{0}^{4}$. The second
striking feature seen in Fig. \ref{fig4} is that the $B(E2)$ values
in $^{109}$I and $^{109}$Te are very sensitive to $\kappa_{\pi
\nu}$. Thus, in panel (b) the transition
$^{109}$I$(11/2_{1}^{+}\rightarrow 7/2_{1}^{+})$ increases rapidly,
from about 500 to more than 1200 $\rm e^{2}\rm fm^{4}$, in the range
of the Figure. Even more striking is what panel (c) shows for the
transition $^{109}$Te$(9/2_{2}^{+}\rightarrow 5/2_{1}^{+})$, for
which the $B(E2)$ value first remains rather constant at about 600
$\rm e^{2}\rm fm^{4}$ to suddenly, at $\kappa_{\pi \nu}=-0.07$
MeV/$r_{0}^{4}$, decrease to reach a vanishing value at $\kappa_{\pi
\nu}=-0.1$ MeV/$r_{0}^{4}$. Finally, in panel (d) the transition
$^{109}$Te$(9/2_{1}^{+}\rightarrow 5/2_{1}^{+})$ increases from zero
to about 300 $\rm{e^{2}fm^{4}}$ when $\kappa_{\pi \nu}$ decreases
from $-0.07$ to $-0.11$ MeV/$r_{0}^{4}$. This may explain why in the
previous shell-model calculation \cite{Te109}, the two
$9/2^+\rightarrow 5/2^+$ transitions are calculated to be inverted.

\begin{figure}
\includegraphics[width = 0.465\textwidth]{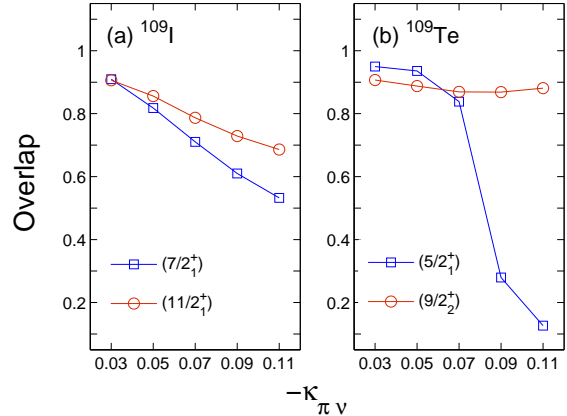}
\caption{(Color online) Absolute values of overlaps for some
low-lying states versus $\kappa_{\pi \nu}$. (a) $^{109}$I with
$\langle \rm ^{109}I(7/2_{1}^{+})|({\pi}g_{7/2})\otimes \rm
^{108}Te(0^{+}_{1})\rangle$ and $\langle \rm
^{109}I(11/2_{1}^{+})|({\pi}g_{7/2})\otimes \rm
^{108}Te(2^{+}_{1})\rangle$, (b) $^{109}$Te with $\langle \rm
^{109}Te(5/2_{1}^{+})|({\nu}d_{5/2})\otimes \rm
^{108}Te(0^{+}_{1})\rangle$ and $\langle \rm
^{109}Te(9/2_{2}^{+})|({\nu}d_{5/2})\otimes \rm
^{108}Te(2^{+}_{1})\rangle$. \label{fig5} }
\end{figure}

This behavior of the $B(E2)$ values in $^{109}$Te, which follows a
rather smooth curve as a function of the proton-neutron interaction
strength, shows a deviation at $\kappa_{\pi \nu}\sim -0.07$
MeV/$r_{0}^{4}$ in panels (c) and (d). There is a peculiarity here,
namely that the state $^{109}$Te$(5/2^+_1)$ is common in the
transitions seen in those panels. We analyze the evolution of the
structures of the states involved in those transitions as a function
of $\kappa_{\pi\nu}$ and found that the dominant NPA configuration
of the state $^{109}$Te$(5/2^+_1)$ changes from
$|(d_{5/2})_{\nu}S_{\nu}^{3}S_{\pi}\rangle$ to
$|(g_{7/2})_{\nu}D_{\nu}S_{\nu}^{2}S_{\pi}\rangle$ when
$\kappa_{\pi\nu}$ varies from $-0.07$ to $-0.09$ MeV/$r_{0}^{4}$.
This change does not occur in the states $^{109}$Te$(9/2_{1}^{+})$
and $^{109}$Te$(9/2_{2}^{+})$.

We complete  the analysis of the weak coupling wave functions by
evaluating, as a function of $\kappa_{\pi\nu}$, their overlaps  with
the states $^{109}$I$(7/2_{1}^{+})$, $^{109}$I$(11/2_{1}^{+})$,
$^{109}$Te$(5/2_{1}^{+})$ and $^{109}$Te$(9/2_{2}^{+})$. These
overlaps are shown in Fig. \ref{fig5}. One sees that the overlaps of
$^{109}$I in Fig. \ref{fig5}(a) decrease with increasing
$-\kappa_{\pi \nu}$, and the most rapid changes occur for the state
$7/2_{1}^{+}$. Instead, the weak coupling description of the state
$^{109}$Te$(9/2_{2}^{+})$ in Fig. \ref{fig5}(b), is practically
independent of $\kappa_{\pi \nu}$. But the most striking feature in
this Figure is the very sharp change of the overlap corresponding to
the state $5/2_{1}^{+}$ when $\kappa_{\pi \nu}$ varies from $-0.07$
to $-0.09$ MeV/$r_{0}^{4}$. As mentioned above, this abrupt change
is a consequence of the evolution of the NPA wave function, which at
that value ($-0.09$ MeV/$r_{0}^{4}$) of the strength the NPA
configuration $|(g_{7/2})_{\nu}D_{\nu}S_{\nu}^{2}S_{\pi}\rangle$
becomes dominant. This analysis shows that the value of the
proton-neutron interaction strength $\kappa_{\pi \nu}$ is weak.
Perhaps most important is that with the weak strength, the
theoretical $B(E2,11/2_{1}^{+}\rightarrow 7/2_{1}^{+})$ value in
$^{109}$I and $B(E2,9/2_{2}^{+}\rightarrow 5/2_{1}^{+})$ value in
$^{109}$Te acquire the experimental value (within the experimental
error) of 762 and 595 $\rm e^2 fm^4$, as shown in Fig. \ref{fig5}
and Table \ref{table:2}.

\section{Summary}

In this paper we have calculated the low-lying level schemes and
electromagnetic transition properties of the nuclei $^{108}$Te,
$^{109}$Te and $^{109}$I within the nucleon pair approximation (NPA)
of the shell model. We extract from the NPA wave functions the
probabilities that the low-lying bands in $^{109}$I and $^{109}$Te
could be interpreted in terms of the weak coupling between the
collective even-even core $^{108}$Te and the unpaired particle. We
thus find that that is indeed the case, as shown in Table
\ref{table:3}. This is consistent with the conjectures put forward
in Refs. \cite{I109-2,Te109}.

We probe the weak coupling pictures of the states by investigating
the corresponding $B(E2)$ values, and find that the calculated
electromagnetic transitions to the ground states of $^{109}$Te and
$^{109}$I are very sensitive to the residual quadrupole-quadrupole
proton-neutron interaction $\kappa_{\pi \nu}$. By comparing with
experimental data, we conclude that the proton-neutron interaction
is weak and that the states $7/2^{+}_{1}$ and  $11/2^{+}_{1}$ in
$^{109}$I as well as the states $5/2^{+}_{1}$ and $9/2^{+}_{2}$ in
$^{109}$Te are well described by the weak coupling scheme.

We tabulate our calculated $B(E2)$, $B(M1)$ and $\mu$ values for
some low-lying states. Except for the transition
$B(E2,9/2_{1}^{+}\rightarrow 5/2_{2}^{+})$, our results agree very
well with available experimental data, thus confirming the
experimental order of the first two excited $9/2^{+}$ states in
$^{109}$Te.

The overall agreement between the calculations and experiments
regarding the $B(E2)$ and $B(M1)$ values, as well as the energy
levels, indicates that the NPA provides an appropriate theoretical
framework to describe low-lying states of these nuclei. Experimental
data are relatively scarce in this region. We therefore believe that our
predictions (e.g., $E2$ and $M1$ transition rates, and magnetic
dipole moments) are useful for future studies of these nuclei.

{\bf Acknowledgements:}

This work was supported by the National Natural Science Foundation
of China (Grant Nos. 11145005, 11225524, 11247241, 11305101 and
11305151), the 973 Program of China (Grant No. 2013CB834401), and
the Shanghai Natural Science Foundation of China (Grant No.
13ZR1419000). C. Qi, R. Liotta and R. Wyss acknowledge the support
of the Swedish Research Council (VR) under grants 621- 2010-4723 and
621-2012-3805. H. Jiang thanks the Shanghai Key Laboratory of
Particle Physics and Cosmology (Grant No. 11DZ2260700) and KTH for
financial support. Discussions with B. Cederwall and T. B$\rm
\ddot{a}$ck are gratefully acknowledged.

\end{document}